\documentclass[12pt,a4paper]{article}
\usepackage{amssymb,amsfonts,amsmath}

\input{tcilatex}
\begin{document}

\title{Context-invariant and local quasi hidden variable (qHV) modelling
versus contextual and nonlocal HV modelling}
\author{Elena R. Loubenets \\
Moscow State Institute of Electronics and Mathematics, \\
Moscow 109028, Russia}
\date{}
\maketitle

\abstract{For all the joint von Neumann measurements on a D-dimensional quantum system, we present the 
specific example of a context-invariant quasi hidden variable (qHV) model, proved in [Loubenets, J. Math. Phys. 56,
032201 (2015)]  to  exist for each Hilbert space. In this model, a quantum observable X  is represented by a variety 
of random variables satisfying the functional condition required in quantum foundations but, in contrast to a contextual 
model, each of  these random variables equivalently models X under all joint von Neumann measurements, regardless
of their contexts.  This, in particular, implies the specific local qHV (LqHV) model for an N-qudit state and allows us to
derive the new exact upper bound on the maximal violation of 2x...x2-setting Bell-type inequalities of any type (either 
on correlation functions or on joint probabilities) under N-partite joint von Neumann measurements on an N-qudit state.
For d=2, this new upper bound coincides with the maximal violation by an N-qubit state of the Mermin-Klyshko
inequality. Based on our results,  we discuss the conceptual and mathematical advantages of context-invariant and
local qHV modelling.}\bigskip

\noindent \textbf{key words: }qHV modelling, nonclassicality, contextuality,
quantum nonlocality, Bell-type inequalities

\bigskip \ 

\section{Introduction}

In quantum theory, the interpretation of von Neumann \cite{1} measurements
via the probability model\footnote{%
In the mathematical physics literature, this probability model is often
named after Kolmogorov. However, in the probability theory literature, the
term "Kolmogorov model" is mostly used \cite{17} for the Kolmogorov
probability axioms \cite{7}. These axioms hold for a measurement of any
nature, in particular, quantum, see also our discussion in \cite{khrenLoub,
13}.} of the classical statistical mechanics, that is, in terms of random
variables and probability measures on a single measurable space\footnote{%
Here, $\Omega $ is a set and $\mathcal{F}_{\Omega }$ is an algebra of
subsets of $\Omega .$} $(\Omega ,\mathcal{F}_{\Omega })$, is referred to as 
\emph{a hidden variable (HV) model. }As it is well known, for all quantum
observables on a Hilbert space of a dimension $\dim \mathcal{H\geq }3,$
there does not \cite{5} exist a noncontextual HV model where random
variables obey the functional subordination inherent to quantum observables.
However, for all observables and states on an arbitrary Hilbert space, there
does exist \cite{holevo} a contextual HV model where this functional
subordination is satisfied. For details and references, see sections 1, 2 in 
\cite{new}.

Under HV modelling \cite{10} of the probabilistic description of a quantum
correlation scenario, noncontextuality is generally interpreted as locality
while contextuality -- as \emph{quantum nonlocality} conjectured by Bell 
\cite{3} due to his result \cite{4} on impossibility of a local HV (LHV)
description of spin measurements of two parties on the two-qubit singlet.
Bell argued \cite{3} that the Einstein-Podolsky-Rosen (EPR) paradox \cite{2}
should be resolved specifically due to the violation of \emph{locality}
under multipartite quantum measurements.

The mathematical result of Bell in \cite{4} can be, however, conditioned by
either of at least two mathematical alternatives: (i) the dependence of a
random variable at one site not only on an observable measured at this site
but also on measurement settings and outcomes at the other sites; (ii)
non-positivity of a scalar measure $\nu $ modelling the singlet state. From
the physical point of view, a choice between these two mathematical
alternatives corresponds to a choice between (i)\textbf{\ }\emph{nonlocality 
}and (ii) \emph{nonclassicality.} The latter results in violation of the
"classical realism" embedded into the probability model of the classical
statistical mechanics just via probability measures. Bell chose the first
alternative and conjectured \cite{4} \emph{quantum nonlocality. }

However, as we stressed in section 3 of \cite{10}, though both, \emph{the
EPR locality and Bell's locality}, correspond to the manifestation of the
physical principle of\emph{\ local action }under nonsignaling multipartite
measurements, but -- the EPR locality, described in \cite{2} as "without in
any way disturbing" systems and measurements at other sites, is a general
concept, not in any way associated with the use of some specific
mathematical formalism, whereas Bell's locality (as it is specified in \cite%
{3,4}) constitutes the manifestation of locality specifically in the HV
frame. As a result, though Bell' locality implies the EPR locality, the
converse is not true -- the EPR locality \emph{does not need} \cite{10} to
imply Bell' locality.

On the other hand, if to view quantum nonlocality only as the manifestation
of quantum entanglement, then it is not clear why, for some nonseparable
quantum states \cite{loubenets2, 10}, all quantum correlation scenarios with 
$S_{n}\leq S_{n}^{(0)}$ settings at each $n$-th site admit the LHV
description while whenever a number of settings $S_{n}>S_{n}^{(0)}$, then
the LHV description of correlation scenarios with such settings is not
possible.

All this\emph{\ questions}\footnote{%
See also in \cite{RL, KH, KH1}.} the conceptual meaning of Bell's "quantum
nonlocality".

Note that, in quantum information theory, a nonlocal nonseparable quantum
state is defined explicitly via nonexistence for all joint von Neumann
measurements on this state of a single LHV\ model formulated in (\cite%
{werner}), therefore, via violation by this state of some Bell-type
inequality\footnote{%
For the general framework on Bell-type inequalities, see \cite{11}.}.

Analyzing the probabilistic description of a general correlation scenario,
we have introduced in \cite{12} a new notion -- the notion of \emph{a local
quasi hidden variable (LqHV) model, }which is specified in terms of "local"
random variables on a measure space $(\Omega ,\mathcal{F}_{\Omega },\nu )$
but, in this triple, a normalized measure $\nu $ can be real-valued. In a
LqHV model, all scenario joint probabilities are reproduced via nonnegative
values of a real-valued measure $\nu $ and all the product expectations --
via the qHV (classical-like) average of the product of the corresponding
"local" random variables.

We\emph{\ }proved \cite{12} that the Hilbert space description of every
quantum correlation scenario admits LqHV modelling. Moreover, we showed \cite%
{13} that the probabilistic description of \emph{every} nonsignaling
correlation scenario does also admit LqHV modelling.

Note that the proved \cite{12, 13} possibility of LqHV modelling of each%
\emph{\ }quantum correlation scenario corresponds just to the second above
alternative -- \emph{nonclassicality }which was, however, disregarded by
Bell.

Based on our results \cite{12, 13} on the probabilistic modelling of
nonsignaling correlation scenarios, we also formulated \cite{13} a new
general probability model, \emph{the quasi-classical probability model, }%
where the measure theory structure inherent to the probability model of the
classical statistical mechanics is preserved but measures $\nu $ on a
measurable space $(\Omega ,\mathcal{F}_{\Omega })$ can be real-valued. In
this new probability model, all observed joint probabilities are reproduced
only via \emph{nonnegative} values of real-valued measures $\nu $. In a
quantum case, we refer to this model as \emph{a quasi hidden variable (qHV)
model. }

Furthermore, we have recently proved by theorem 3 in \cite{new} that, for
each Hilbert space, the Hilbert space description of all the joint von
Neumann measurements admits a \emph{context-invariant} qHV model -- a model
of a completely new type where a quantum observable $X$ can be represented
on a measurable space $(\Omega ,\mathcal{F}_{\Omega })$ by a variety of
random variables satisfying the functional condition required in quantum
foundations, but, in contrast to a contextual\footnote{%
In a contextual model, a quantum observable can be also modelled by a
variety of random variables but which of these random variables represents
an observable $X$ under a joint von Neumann measurement depends \emph{%
specifically on a context} of this joint measurement, i. e. on other
compatible quantum observables measured jointly with $X.$} HV model, each of
these random variables \emph{equivalently} models $X$ under all joint von
Neumann measurements, regardless of their measurement contexts. For $\dim 
\mathcal{H\geq }4,$ the HV version of a context-invariant model for all
quantum observables and states cannot \cite{new} exist.

The proved \cite{new} existence for each Hilbert space of a
context-invariant qHV model, in particular, implies the existence
(proposition 3 in \cite{new}) for each $N$-partite quantum state of a LqHV
model -- the notion introduced in \cite{12}.

In the present paper, we present the specific example (section 2) of a
context-invariant qHV model\emph{\ }\cite{new}\emph{\ }reproducing in
measure theory terms the Hilbert space description of all the joint von
Neumann measurements on a $D$-dimensional quantum system. This, in turn,
implies the specific LqHV model for each $N$-qudit state and allows us to
derive (section 3) the new exact upper bound $\min \{d^{\frac{N-1}{2}%
},3^{N-1}\}$ on the maximal violation of $2\times \cdots \times 2$-setting%
\footnote{%
This notation means \cite{11} that two observables are measured at each of $%
N $ sites.} Bell-type inequalities of any type, either on correlation
functions or on joint probabilities, under $N$-partite joint von Neumann
measurements on an $N$-qudit state. In section 4, we summarize the main
results and stress the mathematical advantages of context-invariant and
local qHV modelling.

\section{Context-invariant qHV modelling}

For the Hilbert space $\mathbb{C}^{D}$, $D\geq 2,$ let us introduce the
specific example of \emph{a context-invariant qHV model }\cite{new}
reproducing the Hilbert space description of all the joint von Neumann
measurements.

Denote by $\mathfrak{X}_{D}$ the set of all quantum observables on $\mathbb{C%
}^{D}$ and by $\Lambda $ the set of all real-valued functions $\lambda :%
\mathfrak{X}_{D}\rightarrow \cup _{X\in \mathfrak{X}_{D}}\mathrm{sp}X$ with
values $\lambda (X)\equiv \lambda _{X}$ in the spectrum $\mathrm{sp}X$ of
the corresponding quantum observable $X$.

Let $\pi _{(X_{1},...,X_{n})}:\Lambda \rightarrow \mathrm{sp}X_{1}\times
\cdots \times \mathrm{sp}X_{n}$ be the canonical projection on $\Lambda :$ 
\begin{eqnarray}
\pi _{(X_{1},...,X_{n})}(\lambda ) &=&\left( \pi _{X_{1}}(\lambda ),...,\pi
_{X_{n}}(\lambda )\right) ,  \label{1} \\
\pi _{X}(\lambda ) &=&\lambda _{X}\in \mathrm{sp}X,  \notag
\end{eqnarray}%
and $\mathcal{A}_{\Lambda }$ be the algebra of all the cylindrical subsets
of the form 
\begin{eqnarray}
\pi _{(X_{1},...,X_{n})}^{-1}(F) &=&\{ \lambda \in \Lambda \mid (\pi
_{X_{1}}(\lambda ),...,\pi _{X_{n}}(\lambda ))\in F\},  \label{2} \\
F &\subseteq &\mathrm{sp}X_{1}\times \cdots \times \mathrm{sp}X_{n},  \notag
\end{eqnarray}%
for all collections $\{X_{1},...,X_{n}\} \subset \mathfrak{X}_{D},n\in 
\mathbb{N},$ of quantum observables on $\mathbb{C}^{D}.$ For a Borel
function $\varphi :\mathbb{R\rightarrow R}$ and an arbitrary quantum
observable $X,$ the random variable $\varphi \circ \pi _{X}\neq \pi
_{\varphi (X)}.$

According to proposition 1 and relation (20) in \cite{new}, to every qudit
state $\rho $ on $\mathbb{C}^{D}$, there corresponds $(\rho \overset{%
\mathfrak{R}}{\mapsto }\mu _{\rho })$ a unique normalized real-valued
measure $\mu _{\rho }:\mathcal{A}_{\Lambda }\rightarrow \mathbb{R},$ $\mu
_{\rho }(\Lambda )=1,$ satisfying the representation%
\begin{eqnarray}
&&\mu _{\rho }(\pi _{(X_{1},...,X_{n})}^{-1}(F))  \label{3} \\
&=&\frac{1}{n!}\dsum \limits_{(x_{1},...,x_{n})\in F}\mathrm{tr}[\rho \{%
\mathrm{P}_{X_{1}}(x_{1})\cdot \ldots \cdot \mathrm{P}_{X_{n}}(x_{n})\}_{%
\mathrm{sym}}]  \notag
\end{eqnarray}%
for all subsets $F\subseteq \mathrm{sp}X_{1}\times \cdots \times \mathrm{sp}%
X_{n}$ and all finite collections $\{X_{1,....,}X_{n}\} \subset \mathfrak{X}%
_{D}$ of quantum observables on $\mathbb{C}^{D}.$ Here, $\mathrm{P}%
_{X_{i}}(\cdot )$ is the spectral (projection-valued) measure of an
observable $X_{i}$ with eigenvalues $\{x_{i}\},$ possibly degenerate, and
the notation $\{ \cdot \}_{sym}$ means the sum arising due to the
symmetrization of the operator product standing in $\{ \cdot \}$ with
respect to all permutations of its factors.

If $\rho _{j}\overset{\mathfrak{R}}{\mapsto }\mu _{\rho _{j}},$ $j=1,...,m,$
then 
\begin{equation}
\sum \alpha _{j}\rho _{j}\overset{\mathfrak{R}}{\mapsto }\sum \alpha _{j}\mu
_{\rho _{j}},\text{ \  \ }\alpha _{j}>0,\  \sum \alpha _{j}=1.
\end{equation}

For mutually commuting quantum observables $X_{1},...,X_{n},$ $n\in \mathbb{N%
},$ representation (\ref{3}) implies 
\begin{eqnarray}
&&\mathrm{tr}[\rho \{ \mathrm{P}_{X_{1}}(B_{1})\cdot \ldots \cdot \mathrm{P}%
_{X_{n}}(B_{n})\}]  \label{4} \\
&=&\mu _{\rho }\left( \pi _{X_{1}}^{-1}(B_{1})\cap \cdots \cap \pi
_{X_{n}}^{-1}(B_{n})\right)  \notag
\end{eqnarray}%
for all subsets $B_{i}\subseteq \mathrm{sp}X_{i},$ $i=1,...,n.$

For each quantum observable $X\in \mathfrak{X}_{D}\mathfrak{,}$ denote by%
\begin{eqnarray}
\lbrack \pi _{X}] &=&\{f_{X,\theta }\mid \theta \in \Theta _{X}\},\text{\  \
where \ }f_{X,\theta _{0}}\equiv \pi _{X},\text{ }  \label{5} \\
f_{X,\theta } &=&\phi _{X}^{(\theta )}\circ \pi _{Y_{\theta }},\text{ \ }%
\phi _{X}^{(\theta )}\circ Y_{\theta }=X,\text{ \ }\phi _{X}^{(\theta )}:%
\mathbb{R}\rightarrow \mathbb{R},\text{ \ }Y_{\theta }\in \mathfrak{X}_{D}, 
\notag
\end{eqnarray}%
the non-empty set of random variables, satisfying the spectral
correspondence rule $f_{X,\theta }(\Lambda )=\mathrm{sp}X$. The index set $%
\Theta _{X}$ depends only on properties of a quantum observable $X.$ If $%
X_{1}\neq X_{2}$, then $\mathfrak{[}\pi _{X_{1}}]\cap \mathfrak{[}\pi
_{X_{2}}]=\varnothing ,$ for the proof, see appendix \textrm{D} in \cite{new}
.

Let $\mathfrak{F}_{\mathfrak{X}_{D}}=\cup _{X\in \mathfrak{X}_{D}}\mathfrak{[%
}\pi _{X}]$ be the union of all the disjoint sets $\mathfrak{[}\pi _{X}],$ $%
X\in \mathfrak{X}_{D},$ of random variables on $\Lambda $ and $\Psi :%
\mathfrak{F}_{\mathfrak{X}_{D}}\rightarrow \mathfrak{X}_{D}$ be the mapping%
\begin{equation}
\Psi (f):=X,\text{ \  \ }f\in \lbrack \pi _{X}]\subset \mathfrak{F}_{%
\mathfrak{X}_{D}},\text{ \  \ }X\in \mathfrak{X}_{D}\mathfrak{,}  \label{6}
\end{equation}%
specifying the correspondence between random variables in the set $\mathfrak{%
F}_{\mathfrak{X}_{D}}$ and quantum observables on $\mathbb{C}^{D}.$ From (%
\ref{5}) it follows that, for each Borel function $\varphi :\mathbb{%
R\rightarrow R}$ and every random variable $f_{X,\theta }\in \lbrack \pi
_{X}]$, we have: 
\begin{eqnarray}
\varphi \circ f_{X,\theta } &=&(\varphi \circ \phi _{X}^{(\theta )})\circ
\pi _{Y_{\theta }}  \label{7} \\
&\in &[\pi _{(\varphi \circ \phi _{X}^{(\theta )})\text{ }\circ Y_{\theta
}}]=[\pi _{\varphi (X)}],  \notag
\end{eqnarray}%
so that $\varphi \circ f_{X,\theta }$ is one of random variables,
representing on $\Lambda $ the quantum observable $\varphi (X)$. Therefore,
the mapping (\ref{6}) satisfies the functional condition (\ref{7}) required
in quantum foundations, see, for example, section 1.4 in \cite{holevo}.

Due to definition (\ref{5}) of sets $[\pi _{X}],$ $X\in \mathfrak{X}_{D},$
and lemma 2 in \cite{new}, the relation 
\begin{eqnarray}
&&\mu _{\rho }\left( \pi _{X_{1}}^{-1}(B_{1})\cap \cdots \cap \pi
_{X_{n}}^{-1}(B_{n})\right)  \label{8} \\
&=&\mu _{\rho }\left( f_{X_{1},\theta _{1}}^{-1}(B_{1})\cap \cdots \cap
f_{X_{n},\theta _{n}}^{-1}(B_{n})\right)  \notag
\end{eqnarray}%
holds for arbitrary random variables $f_{X_{1},\theta _{1}}\in \lbrack \pi
_{X_{1}}],...,f_{X_{n},\theta n}\in \lbrack \pi _{X_{n}}],$ representing on $%
\Lambda $ the corresponding quantum observables.

From (\ref{4}), (\ref{8}) it follows that, for each finite collection $%
\{X_{1},...,X_{n}\},$ $n\in \mathbb{N},$ of mutually commuting observables
on $\mathbb{C}^{D},$ all the von Neumann joint probabilities $\mathrm{tr}%
[\rho \{ \mathrm{P}_{X_{1}}(B_{1})\cdot \ldots \cdot \mathrm{P}%
_{X_{n}}(B_{n})\}],$ $B_{i}\subseteq \mathrm{sp}X_{i},$ admit the
representation%
\begin{eqnarray}
&&\mathrm{tr}[\rho \{ \mathrm{P}_{X_{1}}(B_{1})\cdot \ldots \cdot \mathrm{P}%
_{X_{n}}(B_{n})\}]  \label{9} \\
&=&\mu _{\rho }\left( \pi _{X_{1}}^{-1}(B_{1})\cap \cdots \cap \pi
_{X_{n}}^{-1}(B_{N})\right)  \notag \\
&=&\mu _{\rho }\left( f_{X_{1},\theta _{1}}^{-1}(B_{1})\cap \cdots \cap
f_{X_{n},\theta _{n}}^{-1}(B_{n})\right) ,\text{ \  \  \ }\forall \theta
_{i}\in \Theta _{X_{i}},  \notag
\end{eqnarray}%
which holds for arbitrary random variables $f_{X_{1},\theta _{1}}\in \lbrack
\pi _{X_{1}}],...,f_{X_{n},\theta n}\in \lbrack \pi _{X_{n}}],$ representing
on $\Lambda $ the corresponding quantum observables. For the product
expectations, representation (\ref{9}) immediately implies%
\begin{eqnarray}
&&\mathrm{tr}[\rho (X_{1}\cdot \ldots \cdot X_{n})]  \label{10} \\
&=&\dint \limits_{\Lambda }\pi _{X_{1}}(\lambda )\cdot \ldots \cdot \pi
_{X_{n}}(\lambda )\mu _{\rho }\left( \mathrm{d}\lambda \right)  \notag \\
&=&\dint \limits_{\Lambda }f_{X_{1},\theta _{1}}(\lambda )\cdot \ldots \cdot
f_{X_{n},\theta _{n}}(\lambda )\mu _{\rho }\left( \mathrm{d}\lambda \right) ,%
\text{ \  \ }\forall \theta _{i}\in \Theta _{X_{i}}.  \notag
\end{eqnarray}

Both representations, (\ref{9}) and (\ref{10}), are \emph{context-invariant}
in the sense that, regardless of a context of a joint von Neumann
measurement, into the right-hand sides of these representations, each of
random variables $f_{X,\theta }\in \lbrack \pi _{X}]$ modelling a quantum
observable $X$ can be \emph{equivalently} substituted. Therefore, the set $%
[\pi _{X}],$ defined by relation (\ref{5}), constitutes the class of random
variables equivalently representing $X$ under all joint von Neumann
measurements. The correspondence $[\pi _{X}]\leftrightarrow X,$ specified by
(\ref{1}), (\ref{5}) and (\ref{6}) is one-to-one.

The context-invariant representations (\ref{9}) and (\ref{10}) reproduce the
Hilbert space description of all the joint von Neumann measurements on
qudits in measure theory terms -- i. e. via random variables, satisfying the
functional condition (\ref{7}) generally required in quantum foundations,
and the real-valued normalized measures $\mu _{\rho }$ on the measurable
space $(\Lambda ,\mathcal{A}_{\Lambda }).$ Therefore, these representations
constitute the specific example of \emph{a context-invariant qHV model }%
proved in \cite{new} to exist for each Hilbert space.

\emph{Remark. }Due to representation (\ref{4}), for each state $\rho $ on $%
\mathbb{C}^{D},$ all the elements $\omega _{\rho }(s,U)=\langle s|U\rho
U^{+}|s\rangle =$ $\mathrm{tr}[\rho \{U^{+}|s\rangle \langle s|U\}]\geq 0,$ $%
\sum_{s}\omega _{\rho }(s,U)=1,$ of the quantum state tomogram - the notion
introduced in \cite{manko2}, constitute the corresponding nonnegative values
of the real-valued normalized measure $\mu _{\rho }.$ Here, $U$ is an
unitary operator on $\mathbb{C}^{D}$ and $\{|s\rangle ,s=1,...,D\}$ is an
orthonormal basis in $\mathbb{C}^{D}$.

\section{Local qHV (LqHV) modelling}

Consider now local quasi hidden variable (LqHV) modelling \cite{12} of the
Hilbert space description of $N$-partite joint von Neumann measurements on a
state $\rho _{d,N}$ on $(\mathbb{C}^{d})^{\otimes N}.$

According to proposition 3 in \cite{new}, for each $N$-qudit state $\rho
_{d,N},$ the context-invariant qHV representation (\ref{9}) implies the
following LqHV model:%
\begin{eqnarray}
&&\mathrm{tr}[\rho _{d,N}\text{ }\{ \mathrm{P}_{X_{1}}(B_{1})\otimes \cdots
\otimes \mathrm{P}_{X_{N}}(B_{N})\}]  \label{11} \\
&=&\mu _{\rho _{d,N}}\left( \pi _{\widetilde{X}_{1}}^{-1}(B_{1})\cap \cdots
\cap \pi _{\widetilde{X}_{N}}^{-1}(B_{N})\right)  \notag \\
&=&\dint \limits_{\Lambda }\chi _{\pi _{\widetilde{X}_{1}}^{-1}(B_{1})}(%
\lambda )\cdot \ldots \cdot \chi _{\pi _{\widetilde{X}_{N}}^{-1}(B_{N})}(%
\lambda )\text{ }\mu _{\rho _{d,N}}(\mathrm{d}\lambda ),\text{ \  \ }%
B_{n}\subseteq \mathrm{sp}X_{n},  \notag
\end{eqnarray}%
specified in terms of the measure space $(\Lambda ,\mathcal{A}_{\Lambda
},\mu _{\rho _{d,N}})$ with the real-valued normalized measure $\mu _{\rho
_{d,N}}$ given by relation (\ref{3}) and the "local" random variables 
\begin{eqnarray}
\pi _{\widetilde{X}_{n}}(\lambda ),\text{ \  \ }\widetilde{X}_{n} &=&\mathbb{I%
}_{\mathbb{C}^{d}}\mathbb{\otimes \cdots \otimes \mathbb{I}}_{\mathbb{C}^{d}}%
\mathbb{\otimes }X_{n}\otimes \mathbb{I}_{\mathbb{C}^{d}}\mathbb{\otimes
\cdots \otimes I}_{\mathbb{C}^{d}}, \\
X_{n} &\in &\mathfrak{X}_{d},\text{ \ }n=1,...,N,  \notag
\end{eqnarray}%
where each $\pi _{\widetilde{X}_{n}}$ corresponds by (\ref{1}) to a quantum
observable $X_{n}$ measured at $n$-th site. In (\ref{11}), $\chi
_{A}(\lambda )$ is the indicator function of a set $A\in \mathcal{A}%
_{\Lambda },$ i. e. $\chi _{A}(\lambda )=1,$ if $\lambda \in A,$ and $\chi
_{A}(\lambda )=0,$ if $\lambda \notin A$.

Let us now apply the LqHV representation (\ref{11}) for\ the evaluation
under $N$-partite joint von Neumann measurements on a state $\rho _{d,N}$ of
the maximal quantum violation $\Upsilon _{2\times \cdots \times 2}^{(\rho
_{d,N})}$ of $2\times \cdots \times 2$-setting Bell-type inequalities of any
type, either on correlation functions or on joint probabilities. In
mathematical terms, this state parameter is, in general, defined by Eq. (52)
in \cite{12}.

Denote by $X_{n}^{(i_{n})},$ $i_{n}=1,2,n=1,....,N,$ two quantum
observables, with eigenvalues $\{x_{n}^{(i_{n})}\},$ possibly, degenerate, 
\emph{projectively} measured at each $n$-th of $N$ sites under a $2\times
\cdots \times 2$-setting correlation scenario on a state $\rho _{d,N}.$ Let $%
\tau _{2\times \cdots \times 2}^{(\rho _{d,N})}(\cdot
|X_{1}^{(1)},X_{1}^{(2)},...,X_{N}^{(1)},X_{N}^{(2)})$ be a real-valued
normalized measure in a LqHV model for this scenario. In the LqHV terms, the
expression for $\Upsilon _{2\times \cdots \times 2}^{(\rho _{d,N})}$ follows
from the general relations (40) - (42) in \cite{12} and reads%
\begin{equation}
\Upsilon _{2\times \cdots \times 2}^{(\rho _{d,N})}=\sup_{\substack{ %
X_{n}^{(i_{n})},i_{n}=1,2,  \\ n=1,...,N}}\inf \left \Vert \tau _{2\times
\cdots \times 2}^{(\rho _{d,N})}(\cdot
|X_{1}^{(1)},X_{1}^{(2)},...,X_{N}^{(1)},X_{N}^{(2)})\right \Vert _{var},
\label{gen}
\end{equation}%
where $\left \Vert \tau _{2\times \cdots \times 2}^{(\rho
_{d,N})}\right
\Vert _{var}$ is the total variation norm\footnote{%
On this notion, see, for example, section 3\ in \cite{12}.} of a measure $%
\tau _{2\times \cdots \times 2}^{(\rho _{d,N})}$ and infimum is taken over
all possible LqHV models.

By restricting, in view of (\ref{3}), the measure $\mu _{\rho _{d,N}}$ on $%
\mathcal{A}_{\Lambda }$ to the subalgebra of cylindrical subsets of the form 
$\pi _{(\widetilde{X}_{1}^{(1)},\widetilde{X}_{1}^{(2)},...,\widetilde{X}%
_{N}^{(1)},\widetilde{X}_{N}^{(2)})}^{-1}(F),$ $F\subseteq \Omega ,$ where 
\begin{equation}
\Omega =\mathrm{sp}X_{1}^{(1)}\times \mathrm{sp}X_{1}^{(2)}\times \cdots
\times \mathrm{sp}X_{N}^{(1)}\times \mathrm{sp}X_{N}^{(2)},  \notag
\end{equation}%
and slightly modifying the resulting distribution, for a $2\times \cdots
\times 2$-setting correlation scenario on a state $\rho _{d,N},$ we come due
to (\ref{11}) to the following LqHV model:%
\begin{eqnarray}
&&\mathrm{tr}[\rho _{d,N}\{ \mathrm{P}_{X_{1}^{(i_{1})}}(B_{1}^{(i_{1})})%
\otimes \cdots \otimes \mathrm{P}_{X_{N}^{(i_{N})}}(B_{N}^{(i_{N})})\}]
\label{12} \\
&=&\dsum \limits_{\omega \in \Omega }(\dprod \limits_{n=1,...,N}\chi
_{B_{n}^{(i_{n})}}(x_{n}^{(i_{n})})\text{ })\text{ }\nu _{2\times \cdots
\times 2}^{(\rho _{d,N})}(\omega
|X_{1}^{(1)},X_{1}^{(2)},...,X_{N}^{(1)},X_{N}^{(2)}),  \notag
\end{eqnarray}%
for where $\omega =(x_{1}^{(1)},x_{1}^{(2)},...,x_{N}^{(1)},x_{N}^{(2)}),$ $%
B_{n}^{(i_{n})}\subseteq \mathrm{sp}X_{n}^{(i_{n})},$ \ $i_{n}=1,2,$ and the
real-valued distribution $\nu _{2\times \cdots \times 2}^{(\rho _{d,N})}$ is
specified in Appendix.

By Eq. (\ref{20}), the total variation norm%
\begin{eqnarray}
&&\left \Vert \nu _{2\times \cdots \times 2}^{(\rho _{d,N})}(\cdot
|X_{1}^{(1)},X_{1}^{(2)},...,X_{N}^{(1)},X_{N}^{(2)})\right \Vert _{var} \\
&=&\dsum \limits_{\omega \in \Omega }\left \vert \nu _{2\times \cdots \times
2}^{(\rho _{d,N})}(\omega
|X_{1}^{(1)},X_{1}^{(2)},...,X_{N}^{(1)},X_{N}^{(2)})\right \vert  \notag
\end{eqnarray}%
of the distribution $\nu _{2\times \cdots \times 2}^{(\rho _{d,N})}$ is
upper bounded as 
\begin{equation}
\left \Vert \nu _{2\times \cdots \times 2}^{(\rho _{d,N})}(\cdot
|X_{1}^{(1)},X_{1}^{(2)},...,X_{N}^{(1)},X_{N}^{(2)})\right \Vert _{var}\leq
d^{\frac{N-1}{2}}.  \label{12'}
\end{equation}

From relations (\ref{gen}) and (\ref{12'}) it follows that, for $N$-partite
joint von Neumann measurements on a state $\rho _{d,N},$ \emph{the maximal }$%
2\times \cdots \times 2$-$\emph{setting}$ \emph{Bell violation}%
\begin{equation}
\Upsilon _{2\times \cdots \times 2}^{(\rho _{d,N})}\leq d^{\frac{N-1}{2}}.
\end{equation}%
Taking also into account theorem 4 in \cite{12}, which implies $\Upsilon
_{2\times \cdots \times 2}^{(\rho _{d,N})}\leq 3^{N-1},$ we finally derive 
\begin{equation}
\Upsilon _{2\times \cdots \times 2}^{(\rho _{d,N})}\leq \min \{d^{\frac{N-1}{%
2}},3^{N-1}\}.  \label{13}
\end{equation}%
This new upper bound essentially improves the result following for $S=2$
from the general exact upper bound (62) in \cite{12}.

For $d=2$ and $N=2$, the upper bound in (\ref{13}) is equal to $\sqrt{2}$
and, therefore, coincides with the well-known result on the maximal quantum
violation of Bell-type inequalities on correlation functions and joint
probabilities for two settings and two outcomes per site -- the result
following from the analysis of Fine in \cite{fine} and the Tsirelson maximal
quantum violation bound \cite{tsirelson} for the Clauser-Horne-Shimony-Holt
(CHSH) inequality.

For $d=2$ and an arbitrary $N>2,$ the upper bound in (\ref{13}) is equal to $%
2^{\frac{N-1}{2}}$and, hence, coincides with the maximal violation by an 
\emph{N}-qubit state of the Mermin-Klyshko inequality. Therefore, under $N$%
-partite joint von Neumann measurements on an $N$-qubit state, the
Mermin-Klyshko inequality gives the maximal violation not only among all
correlation $2\times \cdots \times 2$-setting Bell-type inequalities (as it
is proved in \cite{wolf}) but also among $2\times \cdots \times 2$-setting
Bell-type inequalities of any type.

\section{Conclusions}

We have presented the specific example of a context-invariant quasi hidden
variable (qHV) model reproducing due to representations (\ref{9}), (\ref{10}%
) the Hilbert space description of all the joint von Neumann measurements on
qudits via random variables satisfying the functional condition (\ref{7})
required in quantum foundations and the real-valued measures on the
measurable space $(\Lambda ,\mathcal{A}_{\Lambda }).$ In this
context-invariant qHV model, a quantum observable $X$ is represented by the
whole class $[\pi _{X}]$ of random variables, but, in contrast to a
contextual HV model, each of these random variables equivalently models $X$
under all joint von Neumann measurements, regardless of their measurement
contexts. The correspondence $X\leftrightarrow $ $[\pi _{X}],$ specified by
Eqs. (\ref{1}), (\ref{5}) and (\ref{6}), is one-to-one.

For each $N$-qudit state, the context-invariant qHV model (\ref{9}) implies
the specific LqHV model (\ref{12}) which allows us to derive the new exact
upper bound (\ref{13}) on the maximal quantum violation of $2\times \cdots
\times 2$-setting Bell-type inequalities of any type, either on correlation
functions or on joint probabilities, under $N$-partite joint von Neumann
measurements on an $N$ qubit state. Specified for $d=2,$ this new upper
bound coincides with the maximal violation by an $N$ qubit state of the
Mermin-Klyshko inequality. This proves that, under $N$-partite joint von
Neumann measurements on an $N$-qubit state, the Mermin-Klyshko inequality
gives the maximal quantum violation among \emph{all} possible $2\times
\cdots \times 2$-setting Bell-type inequalities, not necessarily the
correlation ones. Note\ that, in the nonlocal HV frame, such quantum
calculations are not possible in principle.

From the conceptual point of view, the new type \cite{new, 12} of
mathematical modelling of joint quantum measurements in measure theory
terms, \emph{context-invariant and local qHV modelling,} corresponds to the
second alternative disregarded by Bell \cite{3, 4} -- \emph{nonclassicality}%
, resulting in violation of the "classical realism" embedded into the
probability model of the classical statistical mechanics via probability
measures. Context-invariant and local qHV modelling\emph{\ }is free from the
conceptual inconsistencies inherent to Bell's concept of "quantum
nonlocality" discussed in Introduction. Moreover, as our results in \cite{12}
and in the present paper demonstrate, this new type of probabilistic
modelling is fruitful for quantum calculations.

\subparagraph{\noindent \textbf{Acknowledgements.}}

I am very grateful to Professor A. Khrennikov for valuable discussions.

\section{Appendix}

Let us first consider the bipartite case $N=2.$ For simplicity of notations,
denote by $X_{i}$, $i=1,2,$ observables measured by Alice and by $Y_{k},$ $%
k=1,2$. measured by Bob. For this case, the values of the real-valued
distribution $\nu _{2\times 2}^{(\rho _{d,2})}(\cdot
|X_{1},X_{2},Y_{1},Y_{2}),$ standing in (\ref{12}), are defined as\medskip 
\begin{align}
& 2\nu _{2\times 2}^{(\rho _{d,2})}(x_{1},x_{2},y_{1},y_{2}\mid
X_{1},X_{2},Y_{1},Y_{2})  \label{14} \\
& =\alpha _{X_{1}}^{(+)}(x_{1}|y_{1},y_{2})\alpha
_{X_{2}}^{(+)}(x_{2}|y_{1},y_{2})\text{ }\mathrm{tr}[\rho _{d,2}\text{ }\{%
\mathbb{I}_{\mathbb{C}^{d}}\text{ }\mathbb{\otimes }\text{ }\mathbb{\{}%
\mathrm{P}_{Y_{1}}(y_{1})\mathrm{P}_{Y_{2}}(y_{2})\}_{\mathrm{sym}}^{(+)}\}]
\notag \\
& -\alpha _{X_{1}}^{(-)}(x_{1}|y_{1},y_{2})\alpha
_{X_{2}}^{(-)}(x_{2}|y_{1},y_{2})\mathrm{tr}[\rho _{d,2}\text{ }\{ \mathbb{I}%
_{\mathbb{C}^{d}}\text{ }\mathbb{\otimes }\text{ }\mathbb{\{}\mathrm{P}%
_{Y_{1}}(y_{1})\mathrm{P}_{Y_{2}}(y_{2})\}_{\mathrm{sym}}^{(-)}\}],  \notag
\end{align}%
\medskip where (i) the notation $Z^{(\pm )}$ means the positive operators,
decomposing a self-adjoint operator $Z=Z^{(+)}-Z^{(-)}$ and satisfying the
relation $Z^{(+)}Z^{(-)}=Z^{(-)}Z^{(+)}=0$, (ii) the probability
distributions $\alpha _{X_{i}}^{(\pm )}(\cdot |y_{1},y_{2}),$ $i=1,2,$ are
defined via the relation 
\begin{eqnarray}
&&\mathrm{tr}[\rho _{d,2}\{ \mathrm{P}_{X_{i}}(x_{i})\otimes \mathbb{\{}%
\mathrm{P}_{Y_{1}}(y_{1})\mathrm{P}_{Y_{2}}(y_{2})\}_{\mathrm{sym}}^{(\pm
)}\}]  \label{15} \\
&=&\alpha _{X_{i}}^{(\pm )}(x_{i}|y_{1},y_{2}))\mathrm{tr}\rho _{d,2}\{%
\mathbb{I}_{\mathbb{C}^{d}}\text{ }\mathbb{\otimes }\text{ }\mathbb{\{}%
\mathrm{P}_{Y_{1}}(y_{1})\mathrm{P}_{Y_{2}}(y_{2})\}_{\mathrm{sym}}^{(\pm
)}\}].  \notag
\end{eqnarray}%
From (\ref{14}) and (\ref{15}) it follows that, for the distribution $\nu
_{2\times 2}^{(\rho _{d,2})},$ the total variation norm%
\begin{eqnarray}
&&\left \Vert \nu _{2\times 2}^{(\rho _{d,2})}(\cdot
|X_{1},X_{2},Y_{1},Y_{2})\right \Vert _{var}  \label{16'} \\
&\equiv &\sum_{\omega \in \Omega }\left \vert \nu _{2\times \cdots \times
2}^{(\rho _{d,N})}(x_{1},x_{2},y_{1},y_{2}|X_{1},X_{2},Y_{1},Y_{2})\right
\vert  \notag \\
&\leq &\frac{1}{2}\left \Vert \dsum \limits_{y_{1},y_{2}}\left \vert \text{ }%
\mathbb{\{}\mathrm{P}_{Y_{1}}(y_{1})\mathrm{P}_{Y_{2}}(y_{2})\}_{\mathrm{sym}%
}\right \vert \right \Vert _{\mathbb{C}^{d}},  \notag
\end{eqnarray}%
where $\left \vert \text{ }\mathbb{\{}\mathrm{P}_{Y_{1}}(y_{1})\mathrm{P}%
_{Y_{2}}(y_{2})\}_{\mathrm{sym}}\right \vert $ is the absolute value operator%
\begin{eqnarray}
&&\left \vert \mathbb{\{}\mathrm{P}_{Y_{1}}(y_{1})\mathrm{P}%
_{Y_{2}}(y_{2})\}_{\mathrm{sym}}\right \vert  \label{17} \\
&=&\mathbb{\{}\mathrm{P}_{Y_{1}}(y_{1})\mathrm{P}_{Y_{2}}(y_{2})\}_{\mathrm{%
sym}}^{(+)}+\mathbb{\{}\mathrm{P}_{Y_{1}}(y_{1})\mathrm{P}_{Y_{2}}(y_{2})\}_{%
\mathrm{sym}}^{(-)}.  \notag
\end{eqnarray}%
Calculating $\left \vert \mathbb{\{}\mathrm{P}_{Y_{1}}(y_{1})\mathrm{P}%
_{Y_{2}}(y_{2})\}_{\mathrm{sym}}\right \vert ,$ we find%
\begin{eqnarray}
&&\dsum \limits_{y_{1},y_{2}}\left \vert \text{ }\mathbb{\{}\mathrm{P}%
_{Y_{1}}(y_{1})\mathrm{P}_{Y_{2}}(y_{2})\}_{\mathrm{sym}}\right \vert
\label{18} \\
&=&\dsum \limits_{k_{1},k_{2}}\left \vert \alpha _{k_{1},k_{2}}\right \vert
\left( |\phi _{Y_{1}}^{(k_{1}}\rangle \langle \phi _{Y_{1}}^{(k_{1}}|\text{ }%
+\text{ }|\phi _{Y_{2}}^{(k_{2}}\rangle \langle \phi
_{Y_{2}}^{(k_{2}}|\right) ,  \notag
\end{eqnarray}%
where $\phi _{Y}^{(k)},$ $k=1,...,d,$ are orthonormal eigenvectors of an
observable $Y$ and $\alpha _{k_{1},k_{2}}=\langle \phi
_{Y_{1}}^{(k_{1})}|\phi _{Y_{2}}^{(k_{2})}\rangle $. Substituting (\ref{18})
into (\ref{16'}) and taking into account that $\dsum
\limits_{k_{i}}\left
\vert \alpha _{k_{1},k_{2}}\right \vert \leq \sqrt{d},$
$i=1,2,$ we finally derive 
\begin{equation}
\left \Vert \nu _{2\times 2}^{(\rho _{d,2})}(\cdot
|X_{1},X_{2},Y_{1},Y_{2})\right \Vert _{var}\leq \sqrt{d}.
\end{equation}

For $N>2,$ the real-valued distribution 
\begin{equation}
\nu _{2\times \cdots \times 2}^{(\rho _{d,N})}(\omega
|X_{1}^{(1)},X_{1}^{(2)},...,X_{N}^{(1)},X_{N}^{(2)})
\end{equation}%
in (\ref{12}) is similar by its construction to distribution (\ref{14}) with
the replacement of $\{ \mathbb{I}_{\mathbb{C}^{d}}\mathbb{\otimes }\frac{1}{2%
}\mathbb{\{}\mathrm{P}_{Y_{1}}(y_{1})\mathrm{P}_{Y_{2}}(y_{2})_{\mathrm{sym}%
}^{(\pm )}\} \}$ by the $N$-partite tensor product of identity operator $%
\mathbb{I}_{\mathbb{C}^{d}}$ and $(N-1)$ factors of the form $\frac{1}{2}%
\mathbb{\{}\mathrm{P}_{X_{n}^{(1)}}(x_{n}^{(1)})\mathrm{P}%
_{X_{n}^{(2)}}(x_{n}^{(2)})\}_{\mathrm{sym}}^{(\pm )}$ at each $n$-th of $%
(N-1)$ sites. As a result,%
\begin{equation}
\left \Vert \nu _{2\times \cdots \times 2}^{(\rho _{d,N})}\right \Vert
_{var}\leq d^{\frac{N-1}{2}}.  \label{20}
\end{equation}%

\end{document}